\documentclass[aps,prl,twocolumn,showpacs,superscriptaddress]{revtex4-1}  % for review and submission

\usepackage{graphicx}  % needed for figures
\usepackage{dcolumn}   % needed for some tables
\usepackage{bm}        % for math
\usepackage{amssymb}  
\usepackage{amsmath}

\begin{document}

% The following information is for internal review, please remove them for submission
\widetext
%\leftline{V.3 To be submitted to PRL}

% the following line is for submission, including submission to the arXiv!!
%\hspace{5.2in} \mbox{Fermilab-Pub-04/xxx-E}

\title{Cold molecular ions on a chip}
%\input author_list.tex       % D0 authors (remove the first 3 lines
                             % of this file prior to submission, they
                             % contain a time stamp for the authorlist)
                             % (includes institutions and visitors)
\author{A. Mokhberi}                     
\author{S. Willitsch}
\email{stefan.willitsch@unibas.ch}
\affiliation{Department of Chemistry, University of Basel, Klingelbergstrasse 80, 4056 Basel, Switzerland}
\date{\today}

\begin{abstract}
We report the sympathetic cooling and Coulomb crystallization of molecular ions above the surface of an ion-trap chip. N$_2^+$ and CaH$^+$ ions were confined in a surface-electrode radiofrequency ion trap and cooled by the interaction with laser-cooled Ca$^{+}$ ions to secular translational temperatures in the millikelvin range. The configuration of trapping potentials generated by the surface electrodes enabled the formation of planar bicomponent Coulomb crystals and the spatial separation of the molecular from the atomic ions on the chip. The structural and thermal properties of the Coulomb crystals were characterized using molecular dynamics simulations. The present study extends chip-based trapping techniques to Coulomb-crystallized molecular ions with potential applications in mass spectrometry, cold chemistry, quantum information science and  spectroscopy.
\end{abstract}

\pacs{37.10.Pq, 37.10.Ty, 37.10.Mn}% Trapping of molecules, Ion trapping, Slowing and cooling of molecules.
\maketitle

%\subsection{\label{sec:level2}Second-level heading: Formatting}
% subsections are not used for PRL papers
%%%%%%%%%%%%%%%%%%%%%%%%%%
%Part0 Introduction

The recent development of miniaturized trapping devices providing tightly confining, highly flexible trapping potentials has paved the way for new schemes for the precise control of neutral atoms and atomic ions. The ability to trap and cool neutral atoms on the surface of microstructured chips has enabled new experiments in the realms of, e. g., quantum optics, quantum interferometry and metrology \citep{AtomChips:2011, Fortgh:2005dm,Treutlein:2010na}. Similarly, the development of radiofrequency (RF) ion-trap chips \cite{Chiaverini:2005, stick06a, Hughes:2011fw} has laid the basis for improved protocols for the manipulation, addressing and shuttling of ions which is of importance for, e.g., quantum information processing  \citep{Kielpinski:2002, ospelkaus11a} and quantum simulation \citep{Clark:2011dj, Szymanski:2012gr, Buluta:2008hs}.

While chip techniques for atomic species are by now well established, their extension to molecules has proven challenging. Translationally cold molecules and molecular ions are currently of great interest for applications in precision spectroscopy \cite{loh13a, hudson11a}, cold chemistry and collision studies \citep{Willitsch2008guide, ospelkaus10b, Hall:2012cw, kirste12a}, quantum optics \citep{Schuster:2011cqed} and quantum information science \citep{DeMille:2002dm, MurPetit:2012gma}. However, significant difficulties arise because the complex molecular energy level structure precludes the implementation of closed optical transitions in most cases. Therefore, standard techniques of atomic physics like laser cooling and fluorescence detection of single particles are at best only applicable to a very restricted number of molecular systems \citep{Shuman:2010hu}. For polar neutral molecules such as CO and Rydberg atoms, the deceleration and trapping on a chip has only recently been achieved using their interaction with time-varying inhomogeneous electric fields generated by microstructured electrodes on a surface \citep{Meek:2009er,Marx:2013gv,hogan12a}. 

Here, we report the generation of Coulomb crystals, i.e., ordered structures of translationally cold and spatially localized molecular ions, above the surface of an ion-trap chip. The molecular ions were cooled sympathetically by the interaction with simultaneously trapped laser-cooled atomic ions \citep{Drewsen:2000, Willitsch:2012gx}. The trapping potential configuration generated by our chip enabled the formation of two- and three-dimensional atomic-molecular bicomponent Coulomb crystals as well as the spatial separation of both species in layers parallel to the chip surface. Our results illustrate the capabilities of chip-based trapping devices for experiments with cold molecular ions with potential applications in mass spectrometry, cold chemistry, quantum information science and spectroscopy.

%Part1-1 The chip 
\begin{figure}[h]
\includegraphics[scale=0.34]{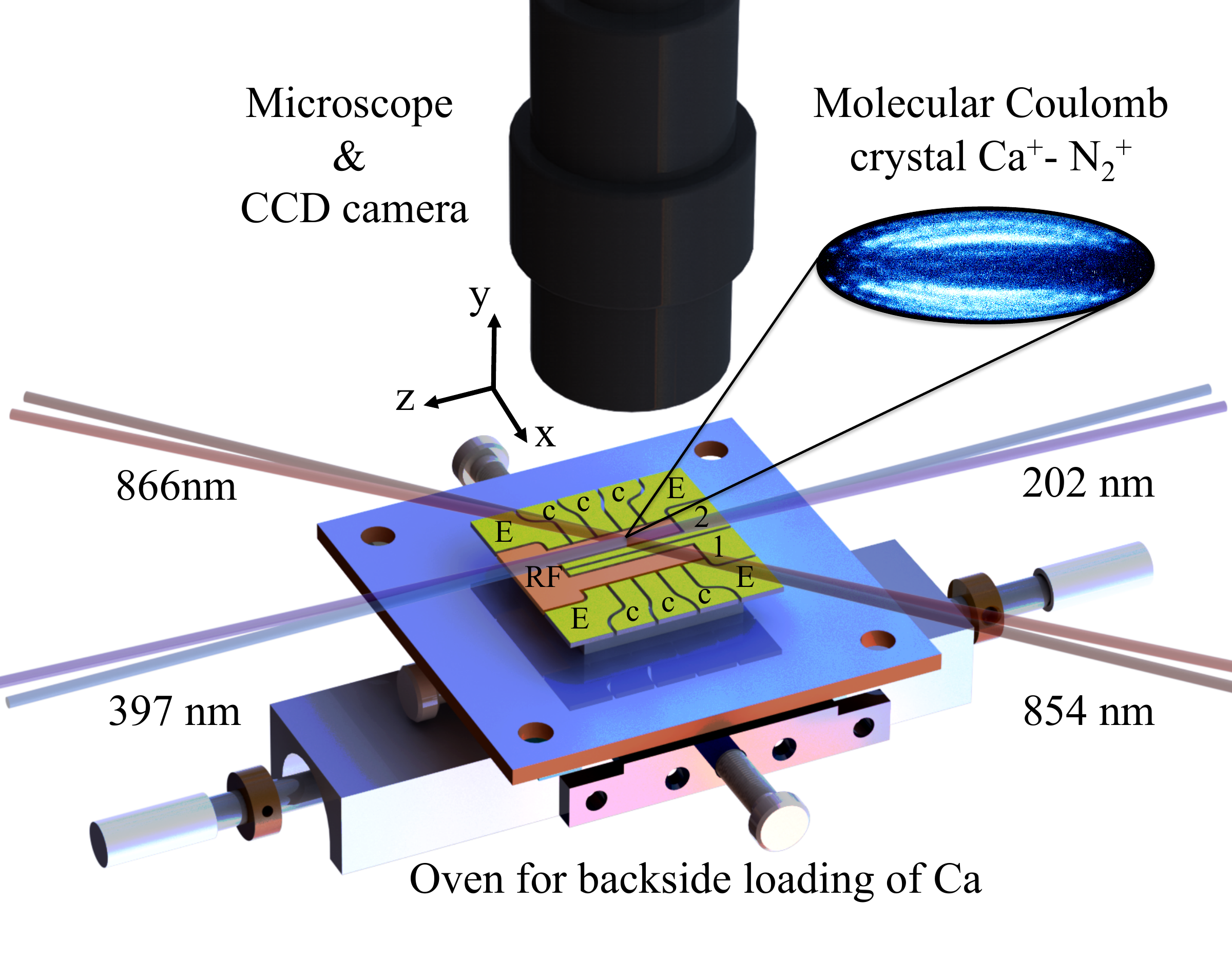}
\caption{\label{fig:experiment} Experimental setup and layout of the surface-electrode ion trap. The electrodes are labeled with $RF$ for the RF electrodes, 1 and 2 for the two central DC electrodes, $E$ for DC endcap electrodes and $c$ for DC control electrodes. The laser beams (397, 866 and 854 nm for the laser cooling of Ca$^{+}$ as well as 202 nm for the photoionization of N$_{2}$ and Ca) were introduced parallel to the chip surface and superimposed in the center of the trap 1.82~mm above the surface. The inset shows a fluorescence image of a Ca$^{+}$-N$_{2}^{+}$ bicomponent Coulomb crystal.}
\end{figure}

%Part1-2 Experimental setup
Inspired by Ref. \citep{Allcock:2010fr}, we adopted a 6-wire surface-electrode (SE) ion trap consisting of a split central electrode, two RF electrodes and two batteries of control and endcap electrodes, see Fig.~\ref{fig:experiment}. The RF electrodes were used to generate time-varying electric fields for the dynamic trapping of the ions in the transverse $(x,y$) directions \cite{Willitsch:2012gx}. All other electrodes were used to apply static (DC) voltages. The static potentials served to confine the ions in the longitudinal $(z)$ direction, to tilt the principal axes of the trap for efficient laser cooling of the atomic ions \cite{Allcock:2010fr}, to compensate excess micromotion \cite{berkeland98a} and to shape the trapping potentials in order to manipulate the structure of the Coulomb crystals \citep{Szymanski:2012gr}. The width of the RF and central electrodes amounted to 2.00~mm and 750~$\mu$m, respectively. The width of the gaps between the electrodes was 300 $\mu$m. The trap electrodes were laser cut out of a 0.50-mm stainless steel foil, electro-polished and coated with a layer of gold with a nominal thickness of 2 $\mu$m. The trap was glued on a ceramic frame separating the electrodes from a printed circuit board containing the trap electronics. This specific trap layout was chosen in order to avoid any dielectric surfaces in the vicinity of the trap center which could cause the buildup of stray charges and patch potentials. 

To load the trap with atomic ions, an atomic beam of neutral calcium was produced by evaporation from a resistively heated stainless steel tube placed underneath the central trap region. The beam passed through a 300 $\mu$m wide slit operating as a skimmer, the ceramic base of the trap and finally the central gap of the chip (between the two central electrodes) to reach the trap center where Ca$^+$ ions were produced by non-resonant photoionization, see Fig.~\ref{fig:experiment}. This backside loading technique has the advantage of avoiding the coating of the chip surface with Ca deposited from the beam  \citep{Amini:2010gw}. Doppler laser cooling of Ca$^{+}$ ions was achieved using three diode laser beams at 397, 866 and 854 nm pumping on the $(4s)^{2} S_{1/2}\rightarrow (4p)^{2} P_{1/2}$, $(3d)^{2}D_{3/2}\rightarrow(4p)^{2} P_{1/2}$ and $(3d)^{2} D_{5/2}\rightarrow(4p)^{2} P_{3/2}$ transitions.

For the present sympathetic cooling experiments, we chose N$_{2}^{+}$ (mass 28 amu) and CaH$^{+}$ (mass 41 amu) as prototypical molecular ions which have a lighter and a heavier mass compared to the laser-cooled Ca$^+$ ions (mass 40 amu), respectively. The N$_{2}^{+}$ ions were produced above the surface of the chip by resonance-enhanced $(2+1)$-photoionization (REMPI) via the $a^{\prime\prime}~^1\Sigma^+_g$ intermediate electronic state of neutral N$_{2}$ \cite{Tong:2010bv} introduced into the vacuum chamber through a leak valve at partial pressure of $<6\times10^{-9}$~mbar. REMPI was carried out using the frequency-tripled output of a Nd:YAG-pumped pulsed dye laser operating at a wavelength of 202 nm and a pulse energy of 80$\mu J$. CaH$^{+}$ ions were produced by chemical reactions of trapped laser-excited Ca$^{+}$ ions with H$_{2}$ molecules leaked into the vacuum chamber. All lasers propagated parallel to the chip surface and were superimposed at the trap center. The fluorescence of the Ca$^+$ ions was spatially resolved by a microscope (magnification $\approx$11.5) and imaged onto a CCD camera. 

%Part2-1 Trapping potentials
The operation of the trap can be understood in terms of an adiabatic approximation for the trapping potentials which assumes that the motion of the ions is slow on the time scale of the RF period \citep{gerlich92a}. Within this approximation, the total effective trapping potential $\Phi_{t}$ experienced by the ions is given by the sum of a time-independent pseudopotential $\Phi_{ps}$ and a static potential $\Phi_{DC}$ generated by the trap electrodes:
\begin{equation}
\begin{split}
\Phi_{t}(x,y,z) &=\Phi_{ps}+\Phi_{DC}\\
&={\ Q^{2} V_{RF}^{2}\over {4 \ M \Omega_{RF}^{2}}} \  { \Vert \nabla \ {\phi_{RF}}\Vert}^2+ Q \sum_{i} V_{i}.\cdot\phi_{i,DC}.
\label{eq:1}
\end{split}
\end{equation}
Here, $Q$ and $M$ are the charge and mass of the trapped ions, respectively, and $V_{RF}$ and $\Omega_{RF}$ denote the RF amplitude and frequency, respectively. The potential terms $\phi_{RF}$ and $\phi_{i,DC}$ are the solution of the Laplace equation for a unit voltage applied to the RF and the i$^{th}$ DC electrode, respectively. V$_{i}$ is the voltage applied on the i$^{th}$ DC electrode.   
 
\begin{figure}
\includegraphics[scale=0.58]{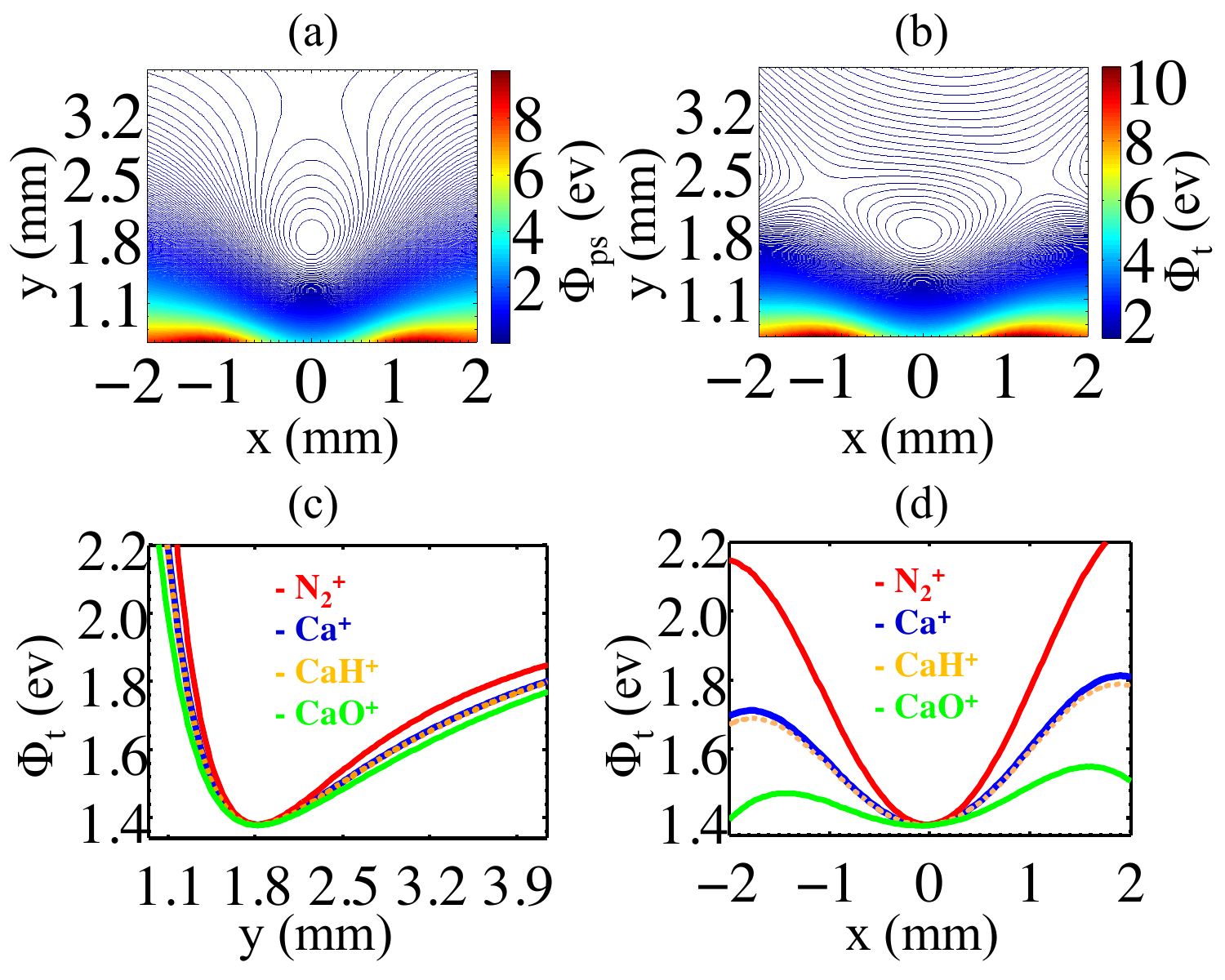}
\caption{\label{fig:potentials} Calculated trapping potentials: (a) Pseudopotential $\Phi_{ps}$ and (b) total potential $\Phi_{t}$ for Ca$^{+}$ ions in the $(x,y)$ plane intersecting the trap center perpendicular to the trap axis. (c),(d) One-dimensional cuts through $\Phi_{t}$ at the trap center for N$_{2}^{+}$, Ca$^{+}$, CaH$^{+}$ and CaO$^{+}$ along the $y$ and $x$ axes, respectively, illustrating the mass dependence of the trapping potential. See Fig. \ref{fig:experiment} for the definition of the coordinate axes. }
\end{figure}

The trapping potentials were calculated by solving the Laplace equation numerically using finite element methods (FEM) as well as using analytical representations of the electrode potentials \citep{schmied10a, SurfacePattern2012}. Results of the FEM calculations are presented in Fig.~\ref{fig:potentials}. The height of trapping, defined as the position of the pseudopotential minimum where the RF field vanishes (the RF null line), was calculated to be 1.82~mm above the surface. The depth of the pseudopotential $\Phi_{ps}$, defined as the potential difference between the minimum and the saddle point above the surface through which ions can escape (see Fig.~\ref{fig:potentials} (a)), was calculated to be 117~meV  for Ca$^{+}$ ions at the radiofrequency voltage V$_{RF}$=495~V and the RF frequency $\Omega_{RF}$=$2\pi\times8.0$~MHz used in the experiments. 

To determine the optimal static voltages applied to the different DC electrodes, the parametrization method of Ref. \citep{Allcock:2010fr} was used. Typical voltages used in the experiments were $\lbrace$central (1,2), control (c), endcap (E) electrodes$\rbrace$=$\lbrace$2 to 2.3, -6 to -6.6, 21.5 to 23$\rbrace$~V. This voltage configuration optimized the total trap depth whilst maintaining the position of the ions at the RF null line to minimize excess micromotion. The application of the static potentials $\Phi_{DC}$ gave rise to saddle points of $\Phi_t$ off-center along the $x$ axis providing additional escape routes of the ions from the trap (see Fig. \ref{fig:potentials} (b)). For Ca$^{+}$, the effective trap depth was limited by the escape of the ions through these saddle points and was calculated to be 113~meV. 

Using the same trapping parameters, the total trapping potential $\Phi_{t}$ has also been calculated for N$_{2}^{+}$, CaH$^{+}$ and CaO$^{+}$ ions. From Eq. \eqref{eq:1}, it can be seen that the pseudopotential $\Phi_{ps}$ is inversely proportional to the ion mass leading to a segregation of the ion species in the Coulomb crystals according to their mass. This situation is completely analogous to the one found in conventional linear RF traps \citep{Willitsch:2012gx}. Fig.~\ref{fig:potentials} (c) and (d) show one-dimensional cuts of $\Phi_t$ at the trap center along the $y$ and $x$-axes for N$_{2}^{+}$, CaH$^{+}$ and CaO$^{+}$ ions in comparison with Ca$^{+}$. For the N$_{2}^{+}$ and CaH$^{+}$ ions used in the experiments, the effective trap depths were calculated to be 261~meV and 104~meV, respectively.

% Part2-2 MDS
To characterize the thermal and structural properties of the Coulomb crystals observed in the experiments, we performed molecular dynamics (MD) simulations \cite{bell09a, Willitsch:2012gx}. Briefly, the classical equations of motion for the laser- and sympathetically cooled ions in the trapping potential were solved numerically. The total force acting on the ions was expressed by
\begin{equation}
\label{eq:2} 
\mathbf{F}_{Total} =\mathbf{F}_{Trap}+\mathbf{F}_{Coulomb}+\mathbf{F}_{LC} +\mathbf{F}_{Heating}+\mathbf{F}_{RP},
\end{equation}
where the terms on the right hand side represent the trapping force, the Coulomb force between ions, the laser cooling force, an effective heating force (reflecting collisions with background gas molecules and imperfections of the setup) and the radiation pressure force generated by unidirectional laser cooling. The laser cooling and radiation pressure forces act only on the atomic ions, while the trapping, Coulomb and heating forces act on both atomic and molecular ions. The ion trajectories obtained from the simulations were used for the reconstruction of the experimental images. A careful comparison of the simulated and experimental images served to determine the number of the ions in the crystal as well as their kinetic energies \citep{bell09a}. To calculate the trapping forces generated by the chip, three-dimensional Taylor expansions were fitted to the numerical electrode potentials. The trap force $\mathbf{F}_{trap}$ in Eq. (\ref{eq:2}) was calculated from analytical gradients of the fitted potential functions. In the simulations, fully time-dependent potentials were used so that effects related the ion micromotion could be characterized.

%Part 3
\begin{figure}
\includegraphics[scale=0.55]{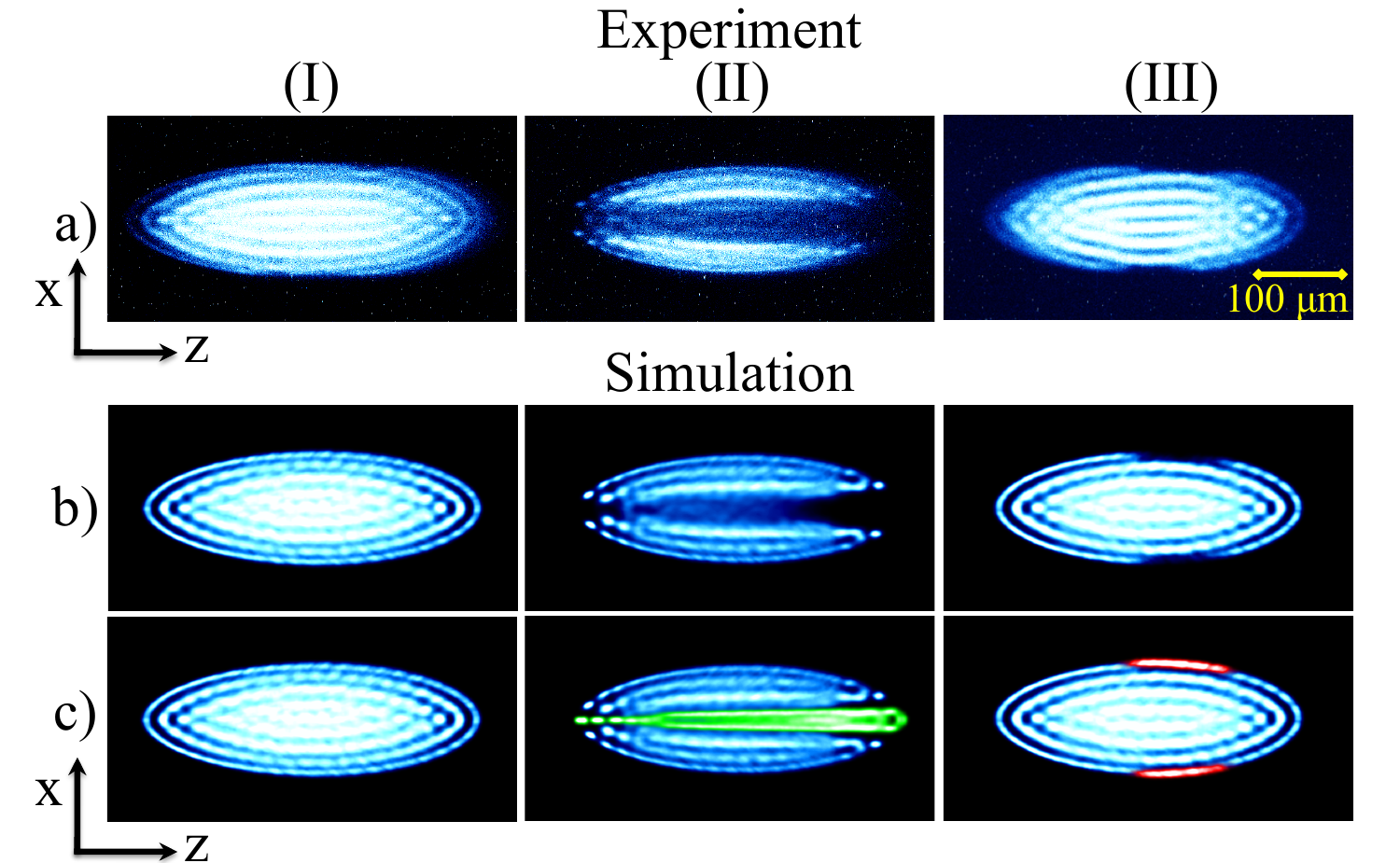}
\caption{\label{fig:LMCC} (a) Experimental and (b,c) simulated false-color laser-cooling fluorescence images of bicomponent Coulomb crystals on the chip. In (c), the sympathetically cooled molecular ions were made visible in the simulations for clarity. Color code: Ca$^+$=blue, N$_2^+$=green and CaH$^+$=red. (I) Pure atomic crystal containing 262 Ca$^{+}$ ions. (II) 50 sympathetically cooled N$_{2}^{+}$  and 180 Ca$^{+}$ ions. (III) 7 sympathetically cooled CaH$^{+}$ and 200 Ca$^{+}$ ions.}
\end{figure}

Fig.~\ref{fig:LMCC} shows (a) experimental and (b, c) simulated images of typical Coulomb crystals imaged in the $(x,z)$ plane parallel to the surface of the chip. The molecular ions do not fluoresce and are only indirectly visible as dark areas appearing in the fluorescence images of the crystals. In Fig.~\ref{fig:LMCC} (c), the sympathetically cooled molecular ions have been made visible in the simulated images for clarity. Column (I) in Fig.~\ref{fig:LMCC} shows a pure atomic Coulomb crystal containing 262 Ca$^{+}$ ions at a mean secular energy E$_{sec}/k_{B}$=23 mK. The secular energy is defined as the thermal energy of the ions without the micromotion. Although the shapes of the crystals projected onto the imaging plane are reminiscent of those obtained in conventional linear RF traps \cite{Willitsch:2012gx}, the Coulomb crystals in the SE trap are not spheroidal owing to the lower symmetry of the trapping potentials (see below and Ref. \cite{Szymanski:2012gr}). This is also reflected by the lack of degeneracy of the trap frequencies in the transverse $(x,y)$ directions. For the present case, the principal trap frequencies have been determined to be $\omega_x/2\pi=171$~kHz, $\omega_y/2\pi=272$~kHz and $\omega_z/2\pi=83$~kHz.

The Ca$^{+}$-N$_{2}^{+}$ bicomponent Coulomb crystal shown in Fig.~\ref{fig:LMCC} (II) (a) was obtained by loading N$_{2}^{+}$ into a Ca$^+$ Coulomb crystal as described above. Based on the MD simulations, we inferred that this  crystal contains 50 N$_{2}^{+}$ ions sympathetically cooled to E$_{sec}/k_{B}$=43 mK in equilibrium with 180 Ca$^{+}$ at E$_{sec}/k_{B}$=33 mK. Since the trapping potential is steeper for lighter ions (see Fig. \ref{fig:potentials} (c, d)), the N$_{2}^{+}$ ions localized closer to the central trap axis than the Ca$^+$ ions. The asymmetric shape of the N$_2^+$ crystal along the $z$-direction was caused by the unidirectional radiation pressure force acting on the Ca$^+$ ions as well as small asymmetries in the potential induced by the geometry of the RF electrodes.

Column (III) of Fig.~\ref{fig:LMCC} shows 7 CaH$^{+}$ ions at E$_{sec}/k_{B}$=28 mK sympathetically cooled by 200 Ca$^{+}$ ions at E$_{sec}/k_{B}$=24 mK.  The heavy CaH$^{+}$ ions localize at the edges of the Ca$^+$ crystal. We found it difficult to sympathetically cool a larger number of CaH$^+$ ions which we attribute to their reduced contact volume with the laser-cooled Ca$^+$ ions at the extremities of the crystal as well as the reduced trap depth for heavier species.
 
%Part 4
\begin{figure}
\includegraphics[scale=0.55]{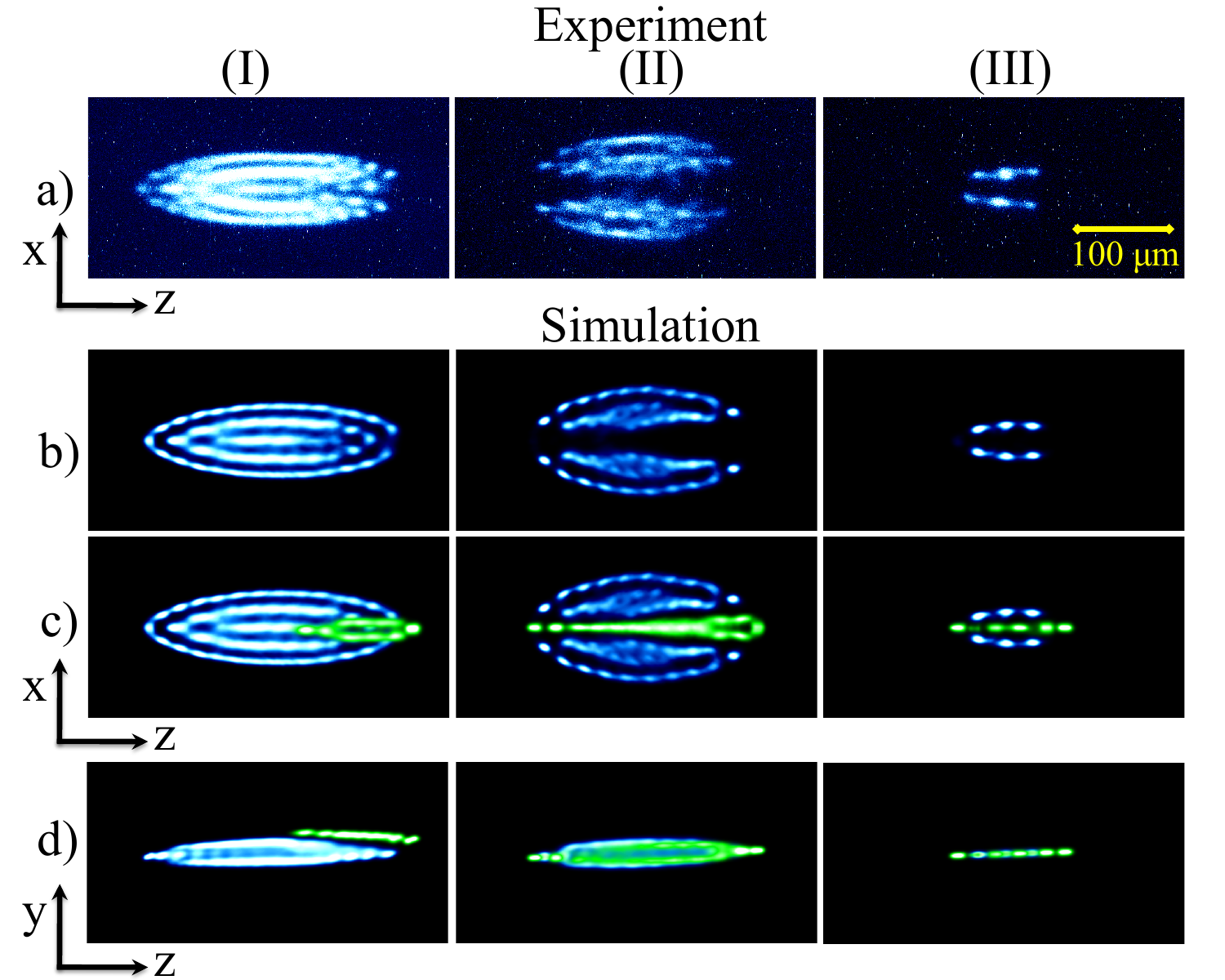}
\caption{\label{fig:planarMCC} Ca$^+$-N$_2^+$ bicomponent Coulomb crystals on the chip at different trapping configurations. (a) Experimental and (b,c,d) simulated false-color laser-cooling fluorescence images. N$_2^+$ ions were made visible in green in the simulations for clarity. In panel (d), simulated images of the side view of the crystals in the y-z plane perpendicular to the chip are shown. (I) 9 N$_{2}^{+}$ ions spatially separated from the Ca$^{+}$ crystal by $\approx$19 $\mu$m in the vertical direction. (II)  18 N$_{2}^{+}$ ions bisecting a near-planar Coulomb crystal with 50 Ca$^{+}$ ions. (III) A string of 5 N$_{2}^{+}$ sympathetically cooled by two strings with 3 Ca$^{+}$ ions each.}
\end{figure}

The flexibility in shaping the trapping potentials on the chip lends itself to a precise manipulation of the crystals to form structures which are challenging to obtain in conventional electrode geometries. In the experiment shown in column (I) of Fig.~\ref{fig:planarMCC}, the molecular and atomic ions were pushed away from the RF null line by decreasing the voltages applied to the central electrodes. Because of the mass dependence of the pseudopotential, the two ion species were spatially separated in the direction perpendicular to the surface such that the N$_{2}^{+}$ ions were located closer to the central trap axis and the Ca$^{+}$ ions further away. The result were two layers of ions with a mean distance of 19$\pm$2 $\mu$m according to the MD simulations as can be seen in the side view of the simulated crystal in Fig.~\ref{fig:planarMCC} (I) (d). 

In the experiment shown in column (II) of Fig. \ref{fig:planarMCC}, a near-planar bicomponent Coulomb crystal was formed with sympathetically cooled N$_2^+$ ions centered along the RF null line splitting the Ca$^+$ crystal in two halves. From the simulated side view of the crystal shown in Fig.~\ref{fig:planarMCC} (II) (d), it can be seen that the crystal has a pancake-like shape and is at maximum only two layers of ions thick.

In the experiment shown in column (III) of Fig. \ref{fig:planarMCC}, a completely planar structure was generated consisting of a string of 5 N$_2^+$ ions sympathetically cooled by two adjacent strings of 3 laser-cooled Ca$^+$ ions each. Such planar bicomponent structures have recently been proposed as a framework for quantum computation and simulation \citep{buluta08b}.

% Summary and Outlook
In summary, we have reported the trapping and sympathetic cooling of molecular ions above the surface of an ion-trap chip. The trapping potentials generated by the electrodes on the chip enabled the formation of two-layer and planar bicomponent Coulomb crystals. Capitalizing on the mass dependence of the pseudopotential, spatially separated layers of atomic and molecular ions could be formed. The flexibility and variability in designing electrode structures in SE traps \citep{Schmied:2009opti, Hughes:2011fw} pave the way for more sophisticated trapping architectures for cold molecular ions with prospects for separating and individually shuttling the different ion species on the chip. Following the recent development of chip-trap devices for atomic ions \cite{Amini:2010gw, Hughes:2011fw}, highly integrated, multifunctional trapping arrays on a single chip can be envisaged which contain dedicated zones for, e.g., the preparation, cooling, spectroscopy, chemistry and mass spectrometry of molecular ions. These developments offer prospects for a significant simplification of the technological overhead for experiments with molecular ions, which have thus far mainly relied on complex and costly guided-ion beam machines \cite{gerlich92a}. 

% Acknowledgement
We gratefully acknowledge the technical support by Dieter Wild and Andreas Tonin. This work was supported by the University of Basel, the COST Action MP1001 "Ion Traps for Tomorrow's Applications" and the Swiss National Science Foundation through the National Centre of Competence in Research "Quantum Science and Technology". 
\bibliographystyle{apsrev4-1}
\bibliography{Mol_ions_on_chip}

\begin{thebibliography}{36}%
\makeatletter
\providecommand \@ifxundefined [1]{%
 \@ifx{#1\undefined}
}%
\providecommand \@ifnum [1]{%
 \ifnum #1\expandafter \@firstoftwo
 \else \expandafter \@secondoftwo
 \fi
}%
\providecommand \@ifx [1]{%
 \ifx #1\expandafter \@firstoftwo
 \else \expandafter \@secondoftwo
 \fi
}%
\providecommand \natexlab [1]{#1}%
\providecommand \enquote  [1]{``#1''}%
\providecommand \bibnamefont  [1]{#1}%
\providecommand \bibfnamefont [1]{#1}%
\providecommand \citenamefont [1]{#1}%
\providecommand \href@noop [0]{\@secondoftwo}%
\providecommand \href [0]{\begingroup \@sanitize@url \@href}%
\providecommand \@href[1]{\@@startlink{#1}\@@href}%
\providecommand \@@href[1]{\endgroup#1\@@endlink}%
\providecommand \@sanitize@url [0]{\catcode `\\12\catcode `\$12\catcode
  `\&12\catcode `\#12\catcode `\^12\catcode `\_12\catcode `\%12\relax}%
\providecommand \@@startlink[1]{}%
\providecommand \@@endlink[0]{}%
\providecommand \url  [0]{\begingroup\@sanitize@url \@url }%
\providecommand \@url [1]{\endgroup\@href {#1}{\urlprefix }}%
\providecommand \urlprefix  [0]{URL }%
\providecommand \Eprint [0]{\href }%
\providecommand \doibase [0]{http://dx.doi.org/}%
\providecommand \selectlanguage [0]{\@gobble}%
\providecommand \bibinfo  [0]{\@secondoftwo}%
\providecommand \bibfield  [0]{\@secondoftwo}%
\providecommand \translation [1]{[#1]}%
\providecommand \BibitemOpen [0]{}%
\providecommand \bibitemStop [0]{}%
\providecommand \bibitemNoStop [0]{.\EOS\space}%
\providecommand \EOS [0]{\spacefactor3000\relax}%
\providecommand \BibitemShut  [1]{\csname bibitem#1\endcsname}%
\let\auto@bib@innerbib\@empty
%</preamble>
\bibitem [{\citenamefont {Reichel}\ and\ \citenamefont
  {Vuleti{\'c}}(2011)}]{AtomChips:2011}%
  \BibitemOpen
  \bibinfo {editor} {\bibfnamefont {J.}~\bibnamefont {Reichel}}\ and\ \bibinfo
  {editor} {\bibfnamefont {V.}~\bibnamefont {Vuleti{\'c}}},\ eds.,\ \href@noop
  {} {\emph {\bibinfo {title} {{Atom chips}}}}\ (\bibinfo  {publisher}
  {WILEY-VCH},\ \bibinfo {address} {Weinheim, Germany},\ \bibinfo {year}
  {2011})\BibitemShut {NoStop}%
\bibitem [{\citenamefont {Fort{\'a}gh}\ and\ \citenamefont
  {Zimmermann}(2005)}]{Fortgh:2005dm}%
  \BibitemOpen
  \bibfield  {author} {\bibinfo {author} {\bibfnamefont {J.}~\bibnamefont
  {Fort{\'a}gh}}\ and\ \bibinfo {author} {\bibfnamefont {C.}~\bibnamefont
  {Zimmermann}},\ }\href@noop {} {\bibfield  {journal} {\bibinfo  {journal}
  {Science}\ }\textbf {\bibinfo {volume} {307}},\ \bibinfo {pages} {860}
  (\bibinfo {year} {2005})}\BibitemShut {NoStop}%
\bibitem [{\citenamefont {Riedel}\ \emph {et~al.}(2010)\citenamefont {Riedel},
  \citenamefont {B{\"o}hi}, \citenamefont {Li}, \citenamefont {H{\"a}nsch},
  \citenamefont {Sinatra},\ and\ \citenamefont {Treutlein}}]{Treutlein:2010na}%
  \BibitemOpen
  \bibfield  {author} {\bibinfo {author} {\bibfnamefont {M.~F.}\ \bibnamefont
  {Riedel}}, \bibinfo {author} {\bibfnamefont {P.}~\bibnamefont {B{\"o}hi}},
  \bibinfo {author} {\bibfnamefont {Y.}~\bibnamefont {Li}}, \bibinfo {author}
  {\bibfnamefont {T.~W.}\ \bibnamefont {H{\"a}nsch}}, \bibinfo {author}
  {\bibfnamefont {A.}~\bibnamefont {Sinatra}}, \ and\ \bibinfo {author}
  {\bibfnamefont {P.}~\bibnamefont {Treutlein}},\ }\href@noop {} {\bibfield
  {journal} {\bibinfo  {journal} {Nature}\ }\textbf {\bibinfo {volume} {464}},\
  \bibinfo {pages} {1170} (\bibinfo {year} {2010})}\BibitemShut {NoStop}%
\bibitem [{\citenamefont {Chiaverini}\ \emph {et~al.}(2005)\citenamefont
  {Chiaverini}, \citenamefont {Blakestad}, \citenamefont {Britton},
  \citenamefont {Jost}, \citenamefont {Langer}, \citenamefont {Leibfried},
  \citenamefont {Ozeri},\ and\ \citenamefont {Wineland}}]{Chiaverini:2005}%
  \BibitemOpen
  \bibfield  {author} {\bibinfo {author} {\bibfnamefont {J.}~\bibnamefont
  {Chiaverini}}, \bibinfo {author} {\bibfnamefont {R.~B.}\ \bibnamefont
  {Blakestad}}, \bibinfo {author} {\bibfnamefont {J.}~\bibnamefont {Britton}},
  \bibinfo {author} {\bibfnamefont {J.~D.}\ \bibnamefont {Jost}}, \bibinfo
  {author} {\bibfnamefont {C.}~\bibnamefont {Langer}}, \bibinfo {author}
  {\bibfnamefont {D.}~\bibnamefont {Leibfried}}, \bibinfo {author}
  {\bibfnamefont {R.}~\bibnamefont {Ozeri}}, \ and\ \bibinfo {author}
  {\bibfnamefont {D.~J.}\ \bibnamefont {Wineland}},\ }\href@noop {} {\bibfield
  {journal} {\bibinfo  {journal} {Quant. Inform. Comp.}\ }\textbf {\bibinfo
  {volume} {5}},\ \bibinfo {pages} {419} (\bibinfo {year} {2005})}\BibitemShut
  {NoStop}%
\bibitem [{\citenamefont {Stick}\ \emph {et~al.}(2006)\citenamefont {Stick},
  \citenamefont {Hensinger}, \citenamefont {Olmschenk}, \citenamefont {Madsen},
  \citenamefont {Schwab},\ and\ \citenamefont {Monroe}}]{stick06a}%
  \BibitemOpen
  \bibfield  {author} {\bibinfo {author} {\bibfnamefont {D.}~\bibnamefont
  {Stick}}, \bibinfo {author} {\bibfnamefont {W.~K.}\ \bibnamefont
  {Hensinger}}, \bibinfo {author} {\bibfnamefont {S.}~\bibnamefont
  {Olmschenk}}, \bibinfo {author} {\bibfnamefont {M.~J.}\ \bibnamefont
  {Madsen}}, \bibinfo {author} {\bibfnamefont {K.}~\bibnamefont {Schwab}}, \
  and\ \bibinfo {author} {\bibfnamefont {C.}~\bibnamefont {Monroe}},\
  }\href@noop {} {\bibfield  {journal} {\bibinfo  {journal} {Nat. Phys.}\
  }\textbf {\bibinfo {volume} {2}},\ \bibinfo {pages} {36} (\bibinfo {year}
  {2006})}\BibitemShut {NoStop}%
\bibitem [{\citenamefont {Hughes}\ \emph {et~al.}(2011)\citenamefont {Hughes},
  \citenamefont {Lekitsch}, \citenamefont {Broersma},\ and\ \citenamefont
  {Hensinger}}]{Hughes:2011fw}%
  \BibitemOpen
  \bibfield  {author} {\bibinfo {author} {\bibfnamefont {M.~D.}\ \bibnamefont
  {Hughes}}, \bibinfo {author} {\bibfnamefont {B.}~\bibnamefont {Lekitsch}},
  \bibinfo {author} {\bibfnamefont {J.~A.}\ \bibnamefont {Broersma}}, \ and\
  \bibinfo {author} {\bibfnamefont {W.~K.}\ \bibnamefont {Hensinger}},\
  }\href@noop {} {\bibfield  {journal} {\bibinfo  {journal} {Contemp. Phys.}\
  }\textbf {\bibinfo {volume} {52}},\ \bibinfo {pages} {505} (\bibinfo {year}
  {2011})}\BibitemShut {NoStop}%
\bibitem [{\citenamefont {Kielpinski}\ \emph {et~al.}(2002)\citenamefont
  {Kielpinski}, \citenamefont {Monroe},\ and\ \citenamefont
  {Wineland}}]{Kielpinski:2002}%
  \BibitemOpen
  \bibfield  {author} {\bibinfo {author} {\bibfnamefont {D.}~\bibnamefont
  {Kielpinski}}, \bibinfo {author} {\bibfnamefont {C.}~\bibnamefont {Monroe}},
  \ and\ \bibinfo {author} {\bibfnamefont {D.~J.}\ \bibnamefont {Wineland}},\
  }\href@noop {} {\bibfield  {journal} {\bibinfo  {journal} {{Nature}
  (London)}\ }\textbf {\bibinfo {volume} {417}},\ \bibinfo {pages} {709}
  (\bibinfo {year} {2002})}\BibitemShut {NoStop}%
\bibitem [{\citenamefont {Ospelkaus}\ \emph {et~al.}(2011)\citenamefont
  {Ospelkaus}, \citenamefont {Warring}, \citenamefont {Colombe}, \citenamefont
  {Brown}, \citenamefont {Amini}, \citenamefont {Leibfried},\ and\
  \citenamefont {Wineland}}]{ospelkaus11a}%
  \BibitemOpen
  \bibfield  {author} {\bibinfo {author} {\bibfnamefont {C.}~\bibnamefont
  {Ospelkaus}}, \bibinfo {author} {\bibfnamefont {U.}~\bibnamefont {Warring}},
  \bibinfo {author} {\bibfnamefont {Y.}~\bibnamefont {Colombe}}, \bibinfo
  {author} {\bibfnamefont {K.~R.}\ \bibnamefont {Brown}}, \bibinfo {author}
  {\bibfnamefont {J.~M.}\ \bibnamefont {Amini}}, \bibinfo {author}
  {\bibfnamefont {D.}~\bibnamefont {Leibfried}}, \ and\ \bibinfo {author}
  {\bibfnamefont {D.~J.}\ \bibnamefont {Wineland}},\ }\href@noop {} {\bibfield
  {journal} {\bibinfo  {journal} {Nature}\ }\textbf {\bibinfo {volume} {476}},\
  \bibinfo {pages} {181} (\bibinfo {year} {2011})}\BibitemShut {NoStop}%
\bibitem [{\citenamefont {Clark}\ \emph {et~al.}(2011)\citenamefont {Clark},
  \citenamefont {Lin}, \citenamefont {Diab},\ and\ \citenamefont
  {Chuang}}]{Clark:2011dj}%
  \BibitemOpen
  \bibfield  {author} {\bibinfo {author} {\bibfnamefont {R.~J.}\ \bibnamefont
  {Clark}}, \bibinfo {author} {\bibfnamefont {Z.}~\bibnamefont {Lin}}, \bibinfo
  {author} {\bibfnamefont {K.~S.}\ \bibnamefont {Diab}}, \ and\ \bibinfo
  {author} {\bibfnamefont {I.~L.}\ \bibnamefont {Chuang}},\ }\href@noop {}
  {\bibfield  {journal} {\bibinfo  {journal} {J. Appl. Phys.}\ }\textbf
  {\bibinfo {volume} {109}},\ \bibinfo {pages} {076103} (\bibinfo {year}
  {2011})}\BibitemShut {NoStop}%
\bibitem [{\citenamefont {Szymanski}\ \emph {et~al.}(2012)\citenamefont
  {Szymanski}, \citenamefont {Dubessy}, \citenamefont {Dubost}, \citenamefont
  {Guibal}, \citenamefont {Likforman},\ and\ \citenamefont
  {Guidoni}}]{Szymanski:2012gr}%
  \BibitemOpen
  \bibfield  {author} {\bibinfo {author} {\bibfnamefont {B.}~\bibnamefont
  {Szymanski}}, \bibinfo {author} {\bibfnamefont {R.}~\bibnamefont {Dubessy}},
  \bibinfo {author} {\bibfnamefont {B.}~\bibnamefont {Dubost}}, \bibinfo
  {author} {\bibfnamefont {S.}~\bibnamefont {Guibal}}, \bibinfo {author}
  {\bibfnamefont {J.~P.}\ \bibnamefont {Likforman}}, \ and\ \bibinfo {author}
  {\bibfnamefont {L.}~\bibnamefont {Guidoni}},\ }\href@noop {} {\bibfield
  {journal} {\bibinfo  {journal} {Appl. Phys. Lett.}\ }\textbf {\bibinfo
  {volume} {100}},\ \bibinfo {pages} {171110} (\bibinfo {year}
  {2012})}\BibitemShut {NoStop}%
\bibitem [{\citenamefont {Buluta}\ \emph {et~al.}(2008)\citenamefont {Buluta},
  \citenamefont {Kitaoka}, \citenamefont {Georgescu},\ and\ \citenamefont
  {Hasegawa}}]{Buluta:2008hs}%
  \BibitemOpen
  \bibfield  {author} {\bibinfo {author} {\bibfnamefont {I.~M.}\ \bibnamefont
  {Buluta}}, \bibinfo {author} {\bibfnamefont {M.}~\bibnamefont {Kitaoka}},
  \bibinfo {author} {\bibfnamefont {S.}~\bibnamefont {Georgescu}}, \ and\
  \bibinfo {author} {\bibfnamefont {S.}~\bibnamefont {Hasegawa}},\ }\href@noop
  {} {\bibfield  {journal} {\bibinfo  {journal} {Phys. Rev. A}\ }\textbf
  {\bibinfo {volume} {77}},\ \bibinfo {pages} {062320} (\bibinfo {year}
  {2008})}\BibitemShut {NoStop}%
\bibitem [{\citenamefont {Loh}\ \emph {et~al.}(2013)\citenamefont {Loh},
  \citenamefont {Cossel}, \citenamefont {Grau}, \citenamefont {Ni},
  \citenamefont {Meyer}, \citenamefont {Bohn}, \citenamefont {Ye},\ and\
  \citenamefont {Cornell}}]{loh13a}%
  \BibitemOpen
  \bibfield  {author} {\bibinfo {author} {\bibfnamefont {H.}~\bibnamefont
  {Loh}}, \bibinfo {author} {\bibfnamefont {K.~C.}\ \bibnamefont {Cossel}},
  \bibinfo {author} {\bibfnamefont {M.~C.}\ \bibnamefont {Grau}}, \bibinfo
  {author} {\bibfnamefont {K.-K.}\ \bibnamefont {Ni}}, \bibinfo {author}
  {\bibfnamefont {E.~R.}\ \bibnamefont {Meyer}}, \bibinfo {author}
  {\bibfnamefont {J.~L.}\ \bibnamefont {Bohn}}, \bibinfo {author}
  {\bibfnamefont {J.}~\bibnamefont {Ye}}, \ and\ \bibinfo {author}
  {\bibfnamefont {E.~A.}\ \bibnamefont {Cornell}},\ }\href@noop {} {\bibfield
  {journal} {\bibinfo  {journal} {Science}\ }\textbf {\bibinfo {volume}
  {342}},\ \bibinfo {pages} {1220} (\bibinfo {year} {2013})}\BibitemShut
  {NoStop}%
\bibitem [{\citenamefont {Hudson}\ \emph {et~al.}(2011)\citenamefont {Hudson},
  \citenamefont {Kara}, \citenamefont {Sallman}, \citenamefont {Sauer},
  \citenamefont {Tarbutt},\ and\ \citenamefont {Hinds}}]{hudson11a}%
  \BibitemOpen
  \bibfield  {author} {\bibinfo {author} {\bibfnamefont {J.~J.}\ \bibnamefont
  {Hudson}}, \bibinfo {author} {\bibfnamefont {D.~M.}\ \bibnamefont {Kara}},
  \bibinfo {author} {\bibfnamefont {I.~J.}\ \bibnamefont {Sallman}}, \bibinfo
  {author} {\bibfnamefont {B.~E.}\ \bibnamefont {Sauer}}, \bibinfo {author}
  {\bibfnamefont {M.~R.}\ \bibnamefont {Tarbutt}}, \ and\ \bibinfo {author}
  {\bibfnamefont {E.~A.}\ \bibnamefont {Hinds}},\ }\href@noop {} {\bibfield
  {journal} {\bibinfo  {journal} {Nature}\ }\textbf {\bibinfo {volume} {473}},\
  \bibinfo {pages} {493} (\bibinfo {year} {2011})}\BibitemShut {NoStop}%
\bibitem [{\citenamefont {Willitsch}\ \emph {et~al.}(2008)\citenamefont
  {Willitsch}, \citenamefont {Bell}, \citenamefont {Gingell}, \citenamefont
  {Procter},\ and\ \citenamefont {Softley}}]{Willitsch2008guide}%
  \BibitemOpen
  \bibfield  {author} {\bibinfo {author} {\bibfnamefont {S.}~\bibnamefont
  {Willitsch}}, \bibinfo {author} {\bibfnamefont {M.~T.}\ \bibnamefont {Bell}},
  \bibinfo {author} {\bibfnamefont {A.~D.}\ \bibnamefont {Gingell}}, \bibinfo
  {author} {\bibfnamefont {S.~R.}\ \bibnamefont {Procter}}, \ and\ \bibinfo
  {author} {\bibfnamefont {T.~P.}\ \bibnamefont {Softley}},\ }\href@noop {}
  {\bibfield  {journal} {\bibinfo  {journal} {Phys. Rev. Lett.}\ }\textbf
  {\bibinfo {volume} {100}},\ \bibinfo {pages} {043203} (\bibinfo {year}
  {2008})}\BibitemShut {NoStop}%
\bibitem [{\citenamefont {Ospelkaus}\ \emph {et~al.}(2010)\citenamefont
  {Ospelkaus}, \citenamefont {Ni}, \citenamefont {Wang}, \citenamefont
  {\mbox{de Miranda}}, \citenamefont {Neyenhuis}, \citenamefont
  {Qu{\'e}m{\'e}ner}, \citenamefont {Julienne}, \citenamefont {Bohn},
  \citenamefont {Jin},\ and\ \citenamefont {Ye}}]{ospelkaus10b}%
  \BibitemOpen
  \bibfield  {author} {\bibinfo {author} {\bibfnamefont {S.}~\bibnamefont
  {Ospelkaus}}, \bibinfo {author} {\bibfnamefont {K.-K.}\ \bibnamefont {Ni}},
  \bibinfo {author} {\bibfnamefont {D.}~\bibnamefont {Wang}}, \bibinfo {author}
  {\bibfnamefont {M.~H.~G.}\ \bibnamefont {\mbox{de Miranda}}}, \bibinfo
  {author} {\bibfnamefont {B.}~\bibnamefont {Neyenhuis}}, \bibinfo {author}
  {\bibfnamefont {G.}~\bibnamefont {Qu{\'e}m{\'e}ner}}, \bibinfo {author}
  {\bibfnamefont {P.~S.}\ \bibnamefont {Julienne}}, \bibinfo {author}
  {\bibfnamefont {J.~L.}\ \bibnamefont {Bohn}}, \bibinfo {author}
  {\bibfnamefont {D.~S.}\ \bibnamefont {Jin}}, \ and\ \bibinfo {author}
  {\bibfnamefont {J.}~\bibnamefont {Ye}},\ }\href@noop {} {\bibfield  {journal}
  {\bibinfo  {journal} {Science}\ }\textbf {\bibinfo {volume} {327}},\ \bibinfo
  {pages} {853} (\bibinfo {year} {2010})}\BibitemShut {NoStop}%
\bibitem [{\citenamefont {Hall}\ and\ \citenamefont
  {Willitsch}(2012)}]{Hall:2012cw}%
  \BibitemOpen
  \bibfield  {author} {\bibinfo {author} {\bibfnamefont {F.~H.~J.}\
  \bibnamefont {Hall}}\ and\ \bibinfo {author} {\bibfnamefont {S.}~\bibnamefont
  {Willitsch}},\ }\href@noop {} {\bibfield  {journal} {\bibinfo  {journal}
  {Phys. Rev. Lett.}\ }\textbf {\bibinfo {volume} {109}},\ \bibinfo {pages}
  {233202} (\bibinfo {year} {2012})}\BibitemShut {NoStop}%
\bibitem [{\citenamefont {Kirste}\ \emph {et~al.}(2012)\citenamefont {Kirste},
  \citenamefont {Wang}, \citenamefont {Schewe}, \citenamefont {Meijer},
  \citenamefont {Liu}, \citenamefont {van~der Avoird}, \citenamefont {Janssen},
  \citenamefont {Gubbels}, \citenamefont {Groenenboom},\ and\ \citenamefont
  {van~de Meerakker}}]{kirste12a}%
  \BibitemOpen
  \bibfield  {author} {\bibinfo {author} {\bibfnamefont {M.}~\bibnamefont
  {Kirste}}, \bibinfo {author} {\bibfnamefont {X.}~\bibnamefont {Wang}},
  \bibinfo {author} {\bibfnamefont {H.~C.}\ \bibnamefont {Schewe}}, \bibinfo
  {author} {\bibfnamefont {G.}~\bibnamefont {Meijer}}, \bibinfo {author}
  {\bibfnamefont {K.}~\bibnamefont {Liu}}, \bibinfo {author} {\bibfnamefont
  {A.}~\bibnamefont {van~der Avoird}}, \bibinfo {author} {\bibfnamefont
  {L.~M.~C.}\ \bibnamefont {Janssen}}, \bibinfo {author} {\bibfnamefont
  {K.~B.}\ \bibnamefont {Gubbels}}, \bibinfo {author} {\bibfnamefont {G.~C.}\
  \bibnamefont {Groenenboom}}, \ and\ \bibinfo {author} {\bibfnamefont
  {S.~Y.~T.}\ \bibnamefont {van~de Meerakker}},\ }\href@noop {} {\bibfield
  {journal} {\bibinfo  {journal} {Science}\ }\textbf {\bibinfo {volume}
  {338}},\ \bibinfo {pages} {1060} (\bibinfo {year} {2012})}\BibitemShut
  {NoStop}%
\bibitem [{\citenamefont {Schuster}\ \emph {et~al.}(2011)\citenamefont
  {Schuster}, \citenamefont {Bishop}, \citenamefont {Chuang}, \citenamefont
  {DeMille},\ and\ \citenamefont {Schoelkopf}}]{Schuster:2011cqed}%
  \BibitemOpen
  \bibfield  {author} {\bibinfo {author} {\bibfnamefont {D.~I.}\ \bibnamefont
  {Schuster}}, \bibinfo {author} {\bibfnamefont {L.~S.}\ \bibnamefont
  {Bishop}}, \bibinfo {author} {\bibfnamefont {I.~L.}\ \bibnamefont {Chuang}},
  \bibinfo {author} {\bibfnamefont {D.}~\bibnamefont {DeMille}}, \ and\
  \bibinfo {author} {\bibfnamefont {R.~J.}\ \bibnamefont {Schoelkopf}},\
  }\href@noop {} {\bibfield  {journal} {\bibinfo  {journal} {Phys. Rev. A}\
  }\textbf {\bibinfo {volume} {83}},\ \bibinfo {pages} {012311} (\bibinfo
  {year} {2011})}\BibitemShut {NoStop}%
\bibitem [{\citenamefont {DeMille}(2002)}]{DeMille:2002dm}%
  \BibitemOpen
  \bibfield  {author} {\bibinfo {author} {\bibfnamefont {D.}~\bibnamefont
  {DeMille}},\ }\href@noop {} {\bibfield  {journal} {\bibinfo  {journal} {Phys.
  Rev. Lett.}\ }\textbf {\bibinfo {volume} {88}},\ \bibinfo {pages} {067901}
  (\bibinfo {year} {2002})}\BibitemShut {NoStop}%
\bibitem [{\citenamefont {Mur-Petit}\ \emph {et~al.}(2012)\citenamefont
  {Mur-Petit}, \citenamefont {Garc{\'\i}a-Ripoll}, \citenamefont
  {P{\'e}rez-R{\'\i}os}, \citenamefont {Campos-Mart{\'\i}nez}, \citenamefont
  {Hern{\'a}ndez},\ and\ \citenamefont {Willitsch}}]{MurPetit:2012gma}%
  \BibitemOpen
  \bibfield  {author} {\bibinfo {author} {\bibfnamefont {J.}~\bibnamefont
  {Mur-Petit}}, \bibinfo {author} {\bibfnamefont {J.~J.}\ \bibnamefont
  {Garc{\'\i}a-Ripoll}}, \bibinfo {author} {\bibfnamefont {J.}~\bibnamefont
  {P{\'e}rez-R{\'\i}os}}, \bibinfo {author} {\bibfnamefont {J.}~\bibnamefont
  {Campos-Mart{\'\i}nez}}, \bibinfo {author} {\bibfnamefont {M.~I.}\
  \bibnamefont {Hern{\'a}ndez}}, \ and\ \bibinfo {author} {\bibfnamefont
  {S.}~\bibnamefont {Willitsch}},\ }\href@noop {} {\bibfield  {journal}
  {\bibinfo  {journal} {Phys. Rev. A}\ }\textbf {\bibinfo {volume} {85}},\
  \bibinfo {pages} {022308} (\bibinfo {year} {2012})}\BibitemShut {NoStop}%
\bibitem [{\citenamefont {Shuman}\ \emph {et~al.}(2010)\citenamefont {Shuman},
  \citenamefont {Barry},\ and\ \citenamefont {DeMille}}]{Shuman:2010hu}%
  \BibitemOpen
  \bibfield  {author} {\bibinfo {author} {\bibfnamefont {E.~S.}\ \bibnamefont
  {Shuman}}, \bibinfo {author} {\bibfnamefont {J.~F.}\ \bibnamefont {Barry}}, \
  and\ \bibinfo {author} {\bibfnamefont {D.}~\bibnamefont {DeMille}},\
  }\href@noop {} {\bibfield  {journal} {\bibinfo  {journal} {Nature}\ }\textbf
  {\bibinfo {volume} {467}},\ \bibinfo {pages} {820} (\bibinfo {year}
  {2010})}\BibitemShut {NoStop}%
\bibitem [{\citenamefont {Meek}\ \emph {et~al.}(2009)\citenamefont {Meek},
  \citenamefont {Conrad},\ and\ \citenamefont {Meijer}}]{Meek:2009er}%
  \BibitemOpen
  \bibfield  {author} {\bibinfo {author} {\bibfnamefont {S.~A.}\ \bibnamefont
  {Meek}}, \bibinfo {author} {\bibfnamefont {H.}~\bibnamefont {Conrad}}, \ and\
  \bibinfo {author} {\bibfnamefont {G.}~\bibnamefont {Meijer}},\ }\href@noop {}
  {\bibfield  {journal} {\bibinfo  {journal} {Science}\ }\textbf {\bibinfo
  {volume} {324}},\ \bibinfo {pages} {1699} (\bibinfo {year}
  {2009})}\BibitemShut {NoStop}%
\bibitem [{\citenamefont {Marx}\ \emph {et~al.}(2013)\citenamefont {Marx},
  \citenamefont {Adu~Smith}, \citenamefont {Abel}, \citenamefont {Zehentbauer},
  \citenamefont {Meijer},\ and\ \citenamefont {Santambrogio}}]{Marx:2013gv}%
  \BibitemOpen
  \bibfield  {author} {\bibinfo {author} {\bibfnamefont {S.}~\bibnamefont
  {Marx}}, \bibinfo {author} {\bibfnamefont {D.}~\bibnamefont {Adu~Smith}},
  \bibinfo {author} {\bibfnamefont {M.~J.}\ \bibnamefont {Abel}}, \bibinfo
  {author} {\bibfnamefont {T.}~\bibnamefont {Zehentbauer}}, \bibinfo {author}
  {\bibfnamefont {G.}~\bibnamefont {Meijer}}, \ and\ \bibinfo {author}
  {\bibfnamefont {G.}~\bibnamefont {Santambrogio}},\ }\href@noop {} {\bibfield
  {journal} {\bibinfo  {journal} {Phys. Rev. Lett.}\ }\textbf {\bibinfo
  {volume} {111}},\ \bibinfo {pages} {243007} (\bibinfo {year}
  {2013})}\BibitemShut {NoStop}%
\bibitem [{\citenamefont {Hogan}\ \emph {et~al.}(2012)\citenamefont {Hogan},
  \citenamefont {Allmendinger}, \citenamefont {Sa\ss{}mannshausen},
  \citenamefont {Schmutz},\ and\ \citenamefont {Merkt}}]{hogan12a}%
  \BibitemOpen
  \bibfield  {author} {\bibinfo {author} {\bibfnamefont {S.~D.}\ \bibnamefont
  {Hogan}}, \bibinfo {author} {\bibfnamefont {P.}~\bibnamefont {Allmendinger}},
  \bibinfo {author} {\bibfnamefont {H.}~\bibnamefont {Sa\ss{}mannshausen}},
  \bibinfo {author} {\bibfnamefont {H.}~\bibnamefont {Schmutz}}, \ and\
  \bibinfo {author} {\bibfnamefont {F.}~\bibnamefont {Merkt}},\ }\href@noop {}
  {\bibfield  {journal} {\bibinfo  {journal} {Phys. Rev. Lett.}\ }\textbf
  {\bibinfo {volume} {108}},\ \bibinfo {pages} {063008} (\bibinfo {year}
  {2012})}\BibitemShut {NoStop}%
\bibitem [{\citenamefont {M\o{}lhave}\ and\ \citenamefont
  {Drewsen}(2000)}]{Drewsen:2000}%
  \BibitemOpen
  \bibfield  {author} {\bibinfo {author} {\bibfnamefont {K.}~\bibnamefont
  {M\o{}lhave}}\ and\ \bibinfo {author} {\bibfnamefont {M.}~\bibnamefont
  {Drewsen}},\ }\href@noop {} {\bibfield  {journal} {\bibinfo  {journal} {Phys.
  Rev. A}\ }\textbf {\bibinfo {volume} {62}},\ \bibinfo {pages} {011401}
  (\bibinfo {year} {2000})}\BibitemShut {NoStop}%
\bibitem [{\citenamefont {Willitsch}(2012)}]{Willitsch:2012gx}%
  \BibitemOpen
  \bibfield  {author} {\bibinfo {author} {\bibfnamefont {S.}~\bibnamefont
  {Willitsch}},\ }\href@noop {} {\bibfield  {journal} {\bibinfo  {journal}
  {Int. Rev. Phys. Chem.}\ }\textbf {\bibinfo {volume} {31}},\ \bibinfo {pages}
  {175} (\bibinfo {year} {2012})}\BibitemShut {NoStop}%
\bibitem [{\citenamefont {Allcock}\ \emph {et~al.}(2010)\citenamefont
  {Allcock}, \citenamefont {Sherman}, \citenamefont {Stacey}, \citenamefont
  {Burrell}, \citenamefont {Curtis}, \citenamefont {Imreh}, \citenamefont
  {Linke}, \citenamefont {Szwer}, \citenamefont {Webster}, \citenamefont
  {Steane},\ and\ \citenamefont {Lucas}}]{Allcock:2010fr}%
  \BibitemOpen
  \bibfield  {author} {\bibinfo {author} {\bibfnamefont {D.~T.~C.}\
  \bibnamefont {Allcock}}, \bibinfo {author} {\bibfnamefont {J.~A.}\
  \bibnamefont {Sherman}}, \bibinfo {author} {\bibfnamefont {D.~N.}\
  \bibnamefont {Stacey}}, \bibinfo {author} {\bibfnamefont {A.~H.}\
  \bibnamefont {Burrell}}, \bibinfo {author} {\bibfnamefont {M.~J.}\
  \bibnamefont {Curtis}}, \bibinfo {author} {\bibfnamefont {G.}~\bibnamefont
  {Imreh}}, \bibinfo {author} {\bibfnamefont {N.~M.}\ \bibnamefont {Linke}},
  \bibinfo {author} {\bibfnamefont {D.~J.}\ \bibnamefont {Szwer}}, \bibinfo
  {author} {\bibfnamefont {S.~C.}\ \bibnamefont {Webster}}, \bibinfo {author}
  {\bibfnamefont {A.~M.}\ \bibnamefont {Steane}}, \ and\ \bibinfo {author}
  {\bibfnamefont {D.~M.}\ \bibnamefont {Lucas}},\ }\href@noop {} {\bibfield
  {journal} {\bibinfo  {journal} {New J. Phys.}\ }\textbf {\bibinfo {volume}
  {12}},\ \bibinfo {pages} {053026} (\bibinfo {year} {2010})}\BibitemShut
  {NoStop}%
\bibitem [{\citenamefont {Berkeland}\ \emph {et~al.}(1998)\citenamefont
  {Berkeland}, \citenamefont {Miller}, \citenamefont {Bergquist}, \citenamefont
  {Itano},\ and\ \citenamefont {Wineland}}]{berkeland98a}%
  \BibitemOpen
  \bibfield  {author} {\bibinfo {author} {\bibfnamefont {D.~J.}\ \bibnamefont
  {Berkeland}}, \bibinfo {author} {\bibfnamefont {J.~D.}\ \bibnamefont
  {Miller}}, \bibinfo {author} {\bibfnamefont {J.~C.}\ \bibnamefont
  {Bergquist}}, \bibinfo {author} {\bibfnamefont {W.~M.}\ \bibnamefont
  {Itano}}, \ and\ \bibinfo {author} {\bibfnamefont {D.~J.}\ \bibnamefont
  {Wineland}},\ }\href@noop {} {\bibfield  {journal} {\bibinfo  {journal} {J.
  Appl. Phys.}\ }\textbf {\bibinfo {volume} {83}},\ \bibinfo {pages} {5025}
  (\bibinfo {year} {1998})}\BibitemShut {NoStop}%
\bibitem [{\citenamefont {Amini}\ \emph {et~al.}(2010)\citenamefont {Amini},
  \citenamefont {Uys}, \citenamefont {Wesenberg}, \citenamefont {Seidelin},
  \citenamefont {Britton}, \citenamefont {Bollinger}, \citenamefont
  {Leibfried}, \citenamefont {Ospelkaus}, \citenamefont {VanDevender},\ and\
  \citenamefont {Wineland}}]{Amini:2010gw}%
  \BibitemOpen
  \bibfield  {author} {\bibinfo {author} {\bibfnamefont {J.~M.}\ \bibnamefont
  {Amini}}, \bibinfo {author} {\bibfnamefont {H.}~\bibnamefont {Uys}}, \bibinfo
  {author} {\bibfnamefont {J.~H.}\ \bibnamefont {Wesenberg}}, \bibinfo {author}
  {\bibfnamefont {S.}~\bibnamefont {Seidelin}}, \bibinfo {author}
  {\bibfnamefont {J.}~\bibnamefont {Britton}}, \bibinfo {author} {\bibfnamefont
  {J.~J.}\ \bibnamefont {Bollinger}}, \bibinfo {author} {\bibfnamefont
  {D.}~\bibnamefont {Leibfried}}, \bibinfo {author} {\bibfnamefont
  {C.}~\bibnamefont {Ospelkaus}}, \bibinfo {author} {\bibfnamefont {A.~P.}\
  \bibnamefont {VanDevender}}, \ and\ \bibinfo {author} {\bibfnamefont {D.~J.}\
  \bibnamefont {Wineland}},\ }\href@noop {} {\bibfield  {journal} {\bibinfo
  {journal} {New J. Phys.}\ }\textbf {\bibinfo {volume} {12}},\ \bibinfo
  {pages} {033031} (\bibinfo {year} {2010})}\BibitemShut {NoStop}%
\bibitem [{\citenamefont {Tong}\ \emph {et~al.}(2010)\citenamefont {Tong},
  \citenamefont {Winney},\ and\ \citenamefont {Willitsch}}]{Tong:2010bv}%
  \BibitemOpen
  \bibfield  {author} {\bibinfo {author} {\bibfnamefont {X.}~\bibnamefont
  {Tong}}, \bibinfo {author} {\bibfnamefont {A.~H.}\ \bibnamefont {Winney}}, \
  and\ \bibinfo {author} {\bibfnamefont {S.}~\bibnamefont {Willitsch}},\
  }\href@noop {} {\bibfield  {journal} {\bibinfo  {journal} {Phys. Rev. Lett.}\
  }\textbf {\bibinfo {volume} {105}},\ \bibinfo {pages} {143001} (\bibinfo
  {year} {2010})}\BibitemShut {NoStop}%
\bibitem [{\citenamefont {Gerlich}(1992)}]{gerlich92a}%
  \BibitemOpen
  \bibfield  {author} {\bibinfo {author} {\bibfnamefont {D.}~\bibnamefont
  {Gerlich}},\ }in\ \href@noop {} {\emph {\bibinfo {booktitle} {{Advances in
  Chemical Physics Series}}}},\ Vol.~\bibinfo {volume} {82},\ \bibinfo {editor}
  {edited by\ \bibinfo {editor} {\bibfnamefont {C.-Y.}\ \bibnamefont {Ng}}\
  and\ \bibinfo {editor} {\bibfnamefont {M.}~\bibnamefont {Baer}}}\ (\bibinfo
  {publisher} {John Wiley \& Sons Inc.},\ \bibinfo {year} {1992})\ p.~\bibinfo
  {pages} {1}\BibitemShut {NoStop}%
\bibitem [{\citenamefont {Schmied}(2010)}]{schmied10a}%
  \BibitemOpen
  \bibfield  {author} {\bibinfo {author} {\bibfnamefont {R.}~\bibnamefont
  {Schmied}},\ }\href@noop {} {\bibfield  {journal} {\bibinfo  {journal} {New
  J. Phys.}\ }\textbf {\bibinfo {volume} {12}},\ \bibinfo {pages} {023038}
  (\bibinfo {year} {2010})}\BibitemShut {NoStop}%
\bibitem [{\citenamefont {Schmied}(2013)}]{SurfacePattern2012}%
  \BibitemOpen
  \bibfield  {author} {\bibinfo {author} {\bibfnamefont {R.}~\bibnamefont
  {Schmied}},\ }\href@noop {} {\enquote {\bibinfo {title} {{SurfacePattern
  software package}},}\ } (\bibinfo {year} {2013})\BibitemShut {NoStop}%
\bibitem [{\citenamefont {Bell}\ \emph {et~al.}(2009)\citenamefont {Bell},
  \citenamefont {Gingell}, \citenamefont {Oldham}, \citenamefont {Softley},\
  and\ \citenamefont {Willitsch}}]{bell09a}%
  \BibitemOpen
  \bibfield  {author} {\bibinfo {author} {\bibfnamefont {M.~T.}\ \bibnamefont
  {Bell}}, \bibinfo {author} {\bibfnamefont {A.~D.}\ \bibnamefont {Gingell}},
  \bibinfo {author} {\bibfnamefont {J.}~\bibnamefont {Oldham}}, \bibinfo
  {author} {\bibfnamefont {T.~P.}\ \bibnamefont {Softley}}, \ and\ \bibinfo
  {author} {\bibfnamefont {S.}~\bibnamefont {Willitsch}},\ }\href@noop {}
  {\bibfield  {journal} {\bibinfo  {journal} {Faraday Discuss. Chem. Soc.}\
  }\textbf {\bibinfo {volume} {142}},\ \bibinfo {pages} {73} (\bibinfo {year}
  {2009})}\BibitemShut {NoStop}%
\bibitem [{\citenamefont {Buluta}\ and\ \citenamefont
  {Hasegawa}(2008)}]{buluta08b}%
  \BibitemOpen
  \bibfield  {author} {\bibinfo {author} {\bibfnamefont {I.~M.}\ \bibnamefont
  {Buluta}}\ and\ \bibinfo {author} {\bibfnamefont {S.}~\bibnamefont
  {Hasegawa}},\ }\href@noop {} {\bibfield  {journal} {\bibinfo  {journal}
  {Phys. Rev. A}\ }\textbf {\bibinfo {volume} {78}},\ \bibinfo {pages} {042340}
  (\bibinfo {year} {2008})}\BibitemShut {NoStop}%
\bibitem [{\citenamefont {Schmied}\ \emph {et~al.}(2009)\citenamefont
  {Schmied}, \citenamefont {Wesenberg},\ and\ \citenamefont
  {Leibfried}}]{Schmied:2009opti}%
  \BibitemOpen
  \bibfield  {author} {\bibinfo {author} {\bibfnamefont {R.}~\bibnamefont
  {Schmied}}, \bibinfo {author} {\bibfnamefont {J.~H.}\ \bibnamefont
  {Wesenberg}}, \ and\ \bibinfo {author} {\bibfnamefont {D.}~\bibnamefont
  {Leibfried}},\ }\href@noop {} {\bibfield  {journal} {\bibinfo  {journal}
  {Phys. Rev. Lett.}\ }\textbf {\bibinfo {volume} {102}},\ \bibinfo {pages}
  {233002} (\bibinfo {year} {2009})}\BibitemShut {NoStop}%
\end{thebibliography}%

\end{document}